# Visualizing the periodic modulation of Cooper pairing in a severely underdoped cuprate

Wei Ruan[1], Xintong Li[1], Cheng Hu[2], Zhenqi Hao[1], Haiwei Li[1], Peng Cai[1], Xingjiang Zhou[2,5], Dung-Hai Lee[3,4], Yayu Wang[1,5,*]

[1]*State Key Laboratory of Low Dimensional Quantum Physics, Department of Physics, Tsinghua University, Beijing 100084, P.R. China*

[2]*Beijing National Laboratory for Condensed Matter Physics, Institute of Physics, Chinese Academy of Sciences, Beijing 100190, P. R. China*

[3]*Department of Physics, University of California at Berkeley, Berkeley, CA 94720.*

[4]*Materials Sciences Division, Lawrence Berkeley National Laboratory, Berkeley, CA 94720.*

[5]*Innovation Center of Quantum Matter, Beijing 100084, P.R. China*

[*] Email: yayuwang@tsinghua.edu.cn

**A major obstacle in understanding the mechanism of Cooper pairing in the cuprates is the existence of various intertwined orders associated with spin, charge, and Cooper pairs[1,2]. Of particular importance is the ubiquitous charge order features that have been observed in a variety of cuprates[2-12], especially in the underdoped regime of the phase diagram. To explain the origin of the charge order and its implication to the superconducting phase, many theoretical models have been proposed, such as charge stripes[13], electronic nematicity[13,14], and Fermi surface instability[6,15]. A highly appealing physical picture is the so-called pair density wave (PDW), a periodic modulation of Cooper paring in space, which may also induce a charge order[2,16-32]. To elucidate the existence and nature of the PDW order, here we use scanning tunneling microscopy (STM) to investigate a severely underdoped $Bi_2Sr_2CaCu_2O_{8+\delta}$, in which superconductivity just emerges on top of a pronounced checkerboard charge order. By analyzing the spatial distribution of the spectral features characteristic of superconductivity, we observe a periodic modulation of both the superconducting coherence peak and gap depth, demonstrating the existence of a density wave order of Cooper pairing. The PDW order has the same spatial periodicity as the charge order, and the amplitudes of the two orders exhibit clear positive correlation. These results shed important new lights on the origin of and interplay between the charge order and Cooper pairing modulation in the cuprates.**

The basic idea of PDW can date back to the original proposal of Fulde-Ferrell[33] and Larkin-Ovchinnikov[34,35], in which the singlet pairing between the Zeeman-split Fermi surfaces will have a finite momentum $Q$, corresponding to periodic spatial modulations. For cuprates, several types of PDW orders have been proposed to be the origin of the charge order[16-30], including unidirectional[16-22], bidirectional[23], and Ampere pairings[24]. More recently, theories involving the intertwinned PDW with *d*-form factor charge order and superconductivity have been extensively studied[25-30], and it is proposed that both orders share a common wavevector.

Although PDW in the cuprates has been predicted for more than 10 years, direct experimental proof of its existence has been lacking. The main challenge is how to probe the spatial distribution of Cooper pairing in the atomic scale. A significant recent progress is the detection of PDW state in optimally doped $Bi_2Sr_2CaCu_2O_{8+\delta}$ (Bi-2212) by using scanning Josephson tunneling microscopy (SJTM)[36]. In this technique, a scanning tip is decorated by a nanometer-sized superconducting (SC) Bi-2212 flake, and the tunneling of Cooper pairs between the tip and sample is utilized to map out the local distribution of Josephson critical current $I_J(r)$. It was found that the superfluid density $\rho_S(r)$ shows a checkerboard pattern indicative of the PDW state, and it shares a common wavevector with the checkerboard charge order. The most probable scenario was proposed to be the coexisting of *d*-wave symmetry superconductivity and *d*-symmetry form factor CDW.

If the PDW state is such a pronounced feature as revealed in the SJTM, one would expect that it should also manifest itself in single particle tunneling spectroscopy probed by plain STM. An obvious strategy is to investigate the spatial distribution of SC gap size, but recent theories show that the SC gap size may not show spatial periodicity even if there is PDW order[24-26,30-32]. Here we propose that there are two other key characteristics associated with superconductivity in a single particle tunneling spectrum. One is the amplitude of the coherence peak near the SC gap edge, which is a direct measure of the SC coherence or superfluid density, as demonstrated by angle-resolved photoemission spectroscopy (ARPES) [37,38]. The other is the depth of the SC gap, which reflects the depletion of low energy quasiparticle density of states (DOS) by Cooper pairing. These two features are directly related to local superconductivity and, as will be shown below, are more sensitive to spatial variations of superconductivity than the gap size.

In order to detect PDW by using STM, we choose a severely underdoped Bi-2212 with $T_c$ ≈ 10 K (hole density $p$ ≈ 0.06), in which superconductivity just emerges by doping the parent Mott insulator. At such low doping, the SC gap and pseudogap are well separated[14,39,40], making it much easier to identify the features associated with superconductivity. Figure 1a displays a large area (600 Å × 600 Å) topographic image acquired on such a Bi-2212 single crystal, where both the atomic structure and supermodulations are clearly resolved. The current map $I(r)$ measured at bias voltage $V$ = 30 mV in the same field of view and the corresponding Fourier transform (FT) are shown in Fig. 1b and c, respectively. The $I(r)$ map measures the spatial distribution of the integrated differential conductance ($dI/dV$) from the Fermi level to 30 mV. It exhibits a pronounced checkerboard pattern with modulation wavevector $Q_{CO}$ = 0.27 ±0.04 ($2\pi/a_0$) along the Cu-Cu bond direction, in excellent agreement with the charge order observed in previous STM results on Bi-2212[4-9]. Individual $dI/dV$ maps at several representative energies are displayed in the supplementary Fig. S1, which clearly demonstrate that the checkerboard pattern is a non-dispersive static charge order. In order to avoid possible set-point effect, all $dI/dV$ curves and mappings are acquired at a set-point bias voltage $V$ = -300 mV, which is properly chosen so that the low-energy particle-hole asymmetric feature only appears at the positive bias[8] (see supplementary session C for a detailed discussion).

Three representative $dI/dV$ curves are plotted in Fig. 1d, taken at locations marked by the corresponding numbered dots in Fig. 1b. They exhibit strong spatial inhomogeneities as reported before[41,42], most likely caused by the random distribution of O defects and Dy dopants. There are two prominent features in these curves: one is the gradual DOS suppression below $\Delta_{PS}$ ~ 100 mV corresponding to the pseudogap, and the other is the small SC gap with size $\Delta_0$

~ 20 mV. For a closer examination of the SC features, in Fig. 1e we zoom into the low bias range of the d$I$/d$V$ spectra, which reveal pronounced variations of the coherence peaks at the SC gap edge (±20 mV). The coherence peaks in the black curve are barely visible, indicating very weak superconductivity. The red curve has clearer features at ±20 mV as two kinks, whereas the blue curve shows very well-defined coherence peaks characteristic of robust superconductivity.

Although it is quite straightforward to see the variations of the coherence peak amplitude from the raw d$I$/d$V$ curves, a more quantitative analysis help reveal its spatial distribution. In order to selectively characterize the coherence peak, we take the minus second derivative $D(V)$ ≡ -(d$I$/d$V$)" = -d$^3I$/d$V^3$ of each d$I$/d$V$ curve. Shown in Fig. 1f are the $D(V)$ plots for the three curves in Fig. 1e, which all exhibit peaks near $V$ = ±20 mV, suggesting that the gap size is almost the same at the three different locations. On the other hand, the peak heights in the $D$ curves are drastically different, confirming the sensitivity of the second derivative to the strength of superconductivity. By definition, the peak height in the $D$ curve directly reflects the local curvature of the d$I$/d$V$ spectrum, or the sharpness of the SC coherence peak, and is insensitive to the d$I$/d$V$ value itself affected by extrinsic factors such as local inhomogeneity, which may superimpose different backgrounds onto the low bias d$I$/d$V$ features at different locations. Therefore the peak in $D$ curve serves as a direct and sensitive indicator of the strength of SC coherence.

Now the validity of the second derivative representation is justified, we can investigate the spatial distribution of SC coherence. Shown in Fig. 2a is a zoomed-in current map in the area indicated by the yellow dashed square in Fig. 1b, and Fig. 2b display the d$I$/d$V$ curves acquired along the yellow arrow in Fig. 2a. They already reveal weak but discernable spatial modulations, where the brighter charge puddles have stronger SC coherence peaks. The second derivative *D* signals are calculated numerically from the d*I*/d*V* data and are shown in Fig. 2c. It becomes evident that the peak heights at *V* = ±20 mV in the *D* curves exhibit periodic spatial patterns in a particle-hole symmetric manner, indicating a periodic modulation of SC coherence. The coherence peaks are the strongest at locations where charge accumulates. These observations are confirmed in another area shown in Fig. 2d, corresponding to the green dashed square area in Fig. 1b. The d*I*/d*V* curves (Fig. 2e) and the calculated *D* curves (Fig. 2f) also show periodic spatial modulations that are commensurate with the charge order. The discernable spatial modulation in the raw d*I*/d*V* curves and the much enhanced periodic patterns in the *D* curves are key characteristics of a PDW closely correlated with the charge order.

To directly visualize the PDW order, the d*I*/d*V* curves are acquired in a dense grid in the whole area in Fig. 1a. The second derivatives are then calculated for each individual curve at bias *V* = 20 mV, and the *D*(*r*, *V* = 20 mV) map is shown in Fig. 3a. This *D*(*r*) map clearly exhibits a checkerboard pattern, which means that the SC coherence, hence the Cooper pair density, has a checkerboard modulation in space. Shown in Fig. 3b is the FT image of the *D*(*r*) map and the PDW wavevector is extracted to be $Q_{PDW} = 0.28 \pm 0.03$ ($2\pi/a_0$), which is the same as the charge order wavevector within joint error bar. To illustrate the relationship between the two orders, in Fig. 3c inset we plot the cross correlation of the *D*(*r*) and *I*(*r*) maps. There is a bright peak located at the center of the correlation map, indicating that the modulations of charge order and PDW are positively correlated and in-phase with each other.

The observation of PDW order from SC coherence is further corroborated by the gap depth analysis. As illustrated in Fig. 4a, the gap depth is defined as the difference between the coherence peak height and the gap bottom in the d*I*/d*V* curve, i.e., *H* = d*I*/d*V*(*V* = 20 mV) – d*I*/d*V*(*V* = 0). This quantity directly reflects the amount of low energy spectral weight that is gapped out by Cooper pairing, thus is another indicator of the local pair density. Figure 4b depicts the gap depth map $H(r)$, extracted from the d$I$/d$V$ curves in the same area, which shows a similar checkerboard pattern as the $D(r)$ map and $I(r)$ map. The modulation wavevector estimated from the FT (Fig. 4c) is $Q_{PDW}$ = 0.28±0.04 ($2\pi/a_0$), consistent with the PDW wavevector extracted from the $D(r)$ map. The cross correlation of $H(r)$ map with the current $I(r)$ map shown in Fig. 4d also displays similar features as in Fig. 3c, demonstrating a positive correlation between the charge and pair density modulation strength. The $D(r)$ and $H(r)$ maps defined at $V$ = -20 mV for the coherence peak at negative bias are shown in supplementary Fig. S2. They exhibit the same checkerboard patterns as that for $V$ = +20 mV, demonstrating that the PDW order is particle-hole symmetric. Furthermore, to remove possible contributions from the charge order signal in our $D$ and $H$ map analysis, we also divide both maps by the integrated total spectral weight between -30 mV and 30 mV. The resulting maps in supplementary Fig. S4 both preserve the checkerboard patterns, which demonstrate that the observed patterns are indeed due to the SC coherence peak variations, rather than the charge order lying within ~ 30 meV around $E_F$.

The STM experiment and data analysis method described here provide a direct way to detect the PDW order from single-particle spectroscopic mapping. Conventional STM data analysis focuses on the d$I$/d$V$ value at a specific bias voltage, whereas here we scrutinize the overall lineshape of a d$I$/d$V$ curve, such as the coherence peak sharpness and gap depth, which are robust indicators of local superconductivity. Several important conclusions can be drawn based on our results. Firstly, we found that the size of the SC gap does not exhibit observable periodic modulations. As shown in Fig. 2c and 2f, the peak positions of the second derivative plots vary for different locations, but in a rather random manner presumably due to local inhomogeneities. In contrast, the SC coherence peak and gap depth show clear checkerboard pattern with the same periodicity as the charge order. Interestingly, this observation is consistent with a recent theory showing that in the presence of intertwined PDW and charge order, the SC gap size barely changes whereas the coherence peak and gap depth exhibit periodic pattern[31,32]. Besides, the charge order mainly gaps out the antinodal Fermi surface[49] while SC occurs near the nodal direction[40], indicating that the SC gap size is unlikely to be affected by the charge order alone. Secondly, the PDW order is observed in a severely underdoped Bi-2212 where superconductivity just emerges. Combining with the previous observation of PDW in optimally doped Bi-2212 by SJTM, the PDW state seems to exist in a wide range of dopings in the cuprate phase diagram. Thirdly, our technique has the advantage of mapping the CDW and PDW orders simultaneously, which allow us to reveal that at least at this particular doping, the pair density and charge density modulations are positively correlated and in-phase with each other. This observation is consistent with the theoretical models where intertwined PDW, $d$-form factor charge order and $d$-wave superconductivity coexist[2,18,22,25-30], but is in contrast to the case of conventional BCS superconductors such as $NbSe_2$, in which the SC coherence is shown to anti-correlate with local charge density wave order due to the competition of the two orders[50].

An important issue that deserves further investigation is whether the PDW is an induced secondary order by the charge order, or the other way around. Our previous STM experiment on lightly doped $Bi_2Sr_2CuO_{6+\delta}$ reveals that the charge order emerges from the parent Mott insulator even before the global superconductivity sets in[10]. This seems to suggest that the holes doped into the Mott insulator first form ordered puddles, upon which the PDW order and superconductivity develop. However, it is also possible that due to strong phase fluctuations at such low doping, the Cooper pairs are strongly localized in space to form a pair crystal, which in return induces a modulated charge distribution. To clarify the exact cause and consequence of these orders, which are critical to the pairing mechanism of cuprates, the narrow regime near the phase boundary when superconductivity just emerges deserves thorough investigations by future STM experiments utilizing the methodology developed here.

**Methods**

High quality Dy doped Bi-2212 single crystals are grown by the floating zone method and are then annealed in vacuum to reach the severely underdoped regime with SC transition temperature ($T_c$) around 10 K. The STM experiments are performed using a low temperature ultrahigh vacuum STM system manufactured by CreaTec. For STM experiments, a single crystal is cleaved in ultrahigh vacuum with pressure better than $10^{-10}$ mbar at $T \sim 77$ K and then immediately transferred to the STM stage sitting at 5 K. Topographic images are scanned using a electrochemically etched tungsten tip which is heated by e-beam and calibrated on an atomically clean Au (111) surface, as described in a previous report [51]. The d$I$/d$V$ spectra are acquired using standard lock-in technique with reference signal frequency $f = 423$ Hz. The d$I$/d$V$ spectra and mappings are all measured at setup bias voltage $V = -300$ mV to avoid set-point effects. The numerical derivative is performed using the Savitzky-Golay method, which fits the experimental data by a 2$^{nd}$-order polynomial and 2$^{nd}$-derivative is obtained from the fit polynomial.

**Data availability**

All raw and derived data used to support the findings of this work are available from the authors on request.

**Acknowledgements:** This work is supported by the NSFC (11190022, 11334010 and 11374335) and MOST of China (2015CB921000), and the Chinese Academy of Sciences (XDB07020300). DHL was supported by the U.S. Department of Energy, Office of Science, Basic Energy Sciences, Materials Sciences and Engineering Division, grant DE-AC02-05CH11231.

**Figure Captions:**

**Figure 1 Topography, charge order, and local electronic structure of underdoped Bi-2212 ($T_c \approx 10$ K).** (a) Topographic image of a 600 Å square area acquired using bias voltage $V = -300$ mV. (b) Tunneling current map $I(r)$ at bias voltage $V = 30$ mV, acquired in the same area in (a). The checkerboard charge order is clearly observed. (c) Fourier transform (FT) of the current map, with modulation wave vector $Q_{CO} = 0.27 \pm 0.04$ $(2\pi/a_0)$. (d) Three typical STS curves taken at the spots indicated by red crosses in (b). Both the pseudogap (~ 100 mV) and SC gap (~ 20 mV) are resolved. The SC gap shows pronounced spatial variations. (e) The corresponding low-bias spectroscopy of the three curves in (d). Vertical offset is used for clarity in both (d) and (e). (f) Minus second derivative of the typical STS curves ($D(V) = -d^3I/dV^3$) with enhanced SC coherence peak feature.

**Figure 2 Periodic modulations of the SC coherence peak feature.** (a) A close-up of the current map $I(r)$ at bias $V = 30$ mV, in the area as indicated by the black dashed square in Fig 1b. (b) Low-bias spectroscopy measured along the yellow arrow shown in (a). (c) $D(V)$ curves obtained from the STS curves in (b), where the periodic modulations of the SC gap are better visualized by the peaks at $\pm 20$ mV. (d)-(f) The same as (a)-(c), but for another area indicated by the white dashed square in Fig. 1b. The STS and $D$ curves are acquired along the yellow arrow in (d).

**Figure 3 Visualization of the pair density wave state by minus second derivative $D(r)$ map.** (a) Spatial distribution of the SC coherence peak feature obtained from the $D(r)$ maps at $V = 20$ mV, which indicates a checkerboard pattern of the Cooper pair density. (b) FT of the $D$ map, with modulation wavevector $Q_{PDW} = 0.28 \pm 0.03$ $(2\pi/a_0)$. (c) Cross-correlation map of the $D(r)$ and $I(r)$ (Fig. 1b) maps.

**Figure 4 Visualization of the pair density wave state by the gap depth map $H(r)$.** (a) Illustration of the gap depth $H$, which is defined by the SC peak height at $V = 20$ mV measured from the zero-bias conductance where the gap bottom sits. (b) The gap depth map $H(r)$ shows similar checkerboard pattern, indicating a spatial modulation of the SC order parameter $\Delta(r)$. (c) FT of the $H(r)$ map, with modulation wavevector $Q_{PDW} = 0.28 \pm 0.04 (2\pi/a_0)$. (d) Cross-correlation map of the $H(r)$ and $I(r)$ maps.

# Figure 1

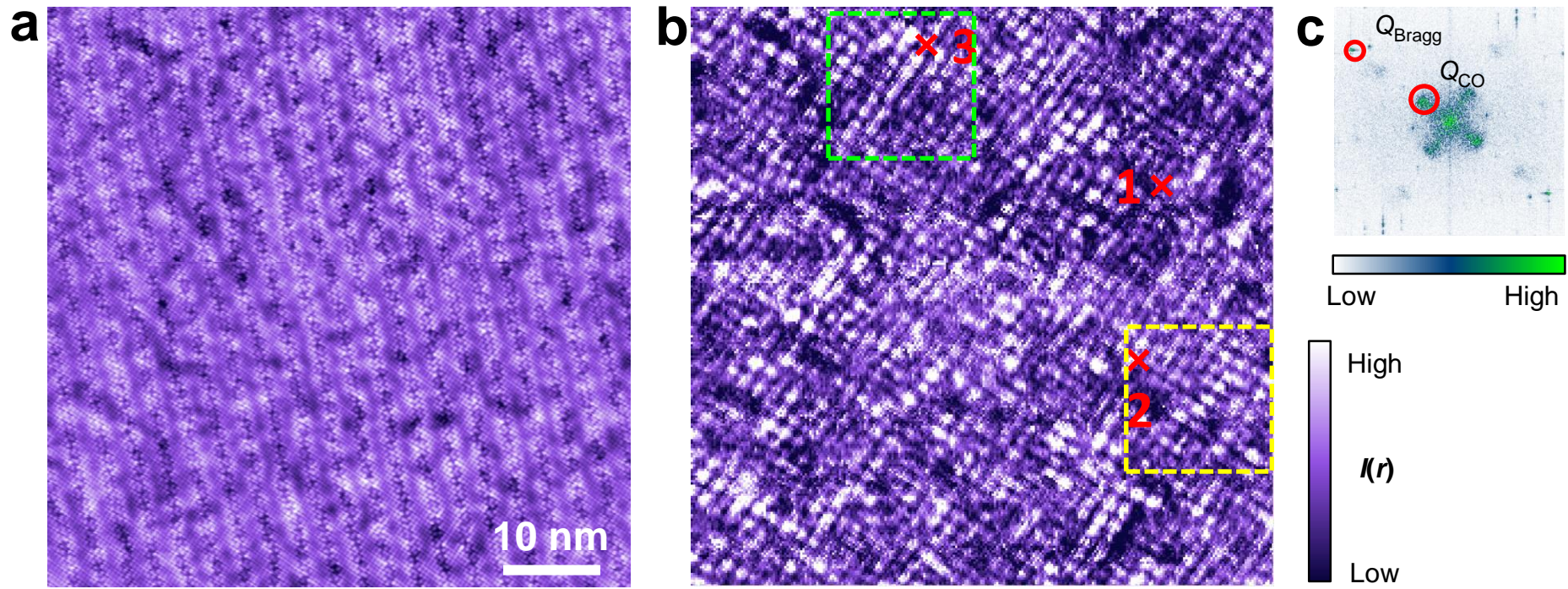

**Figure 2**

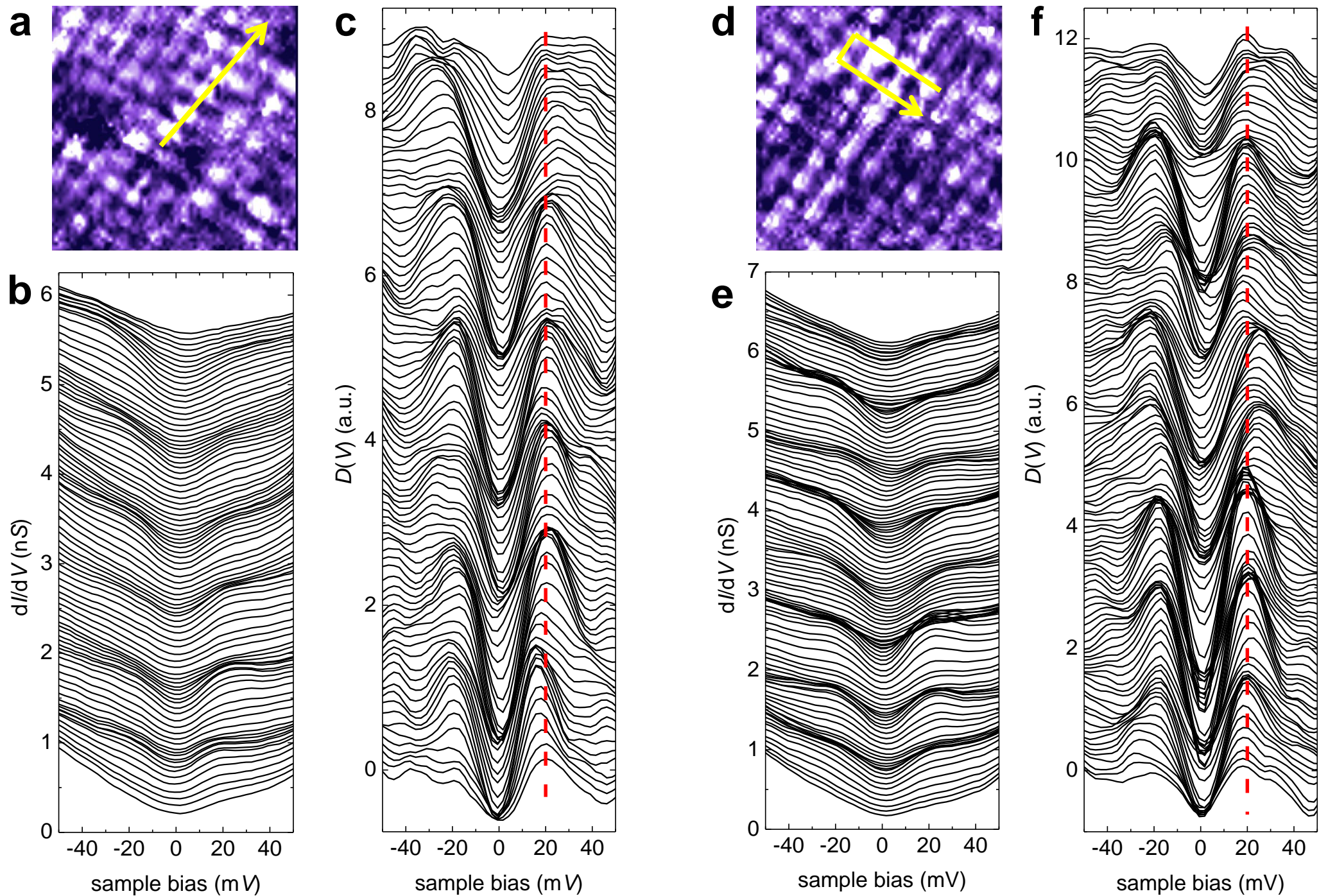

**Figure 3**

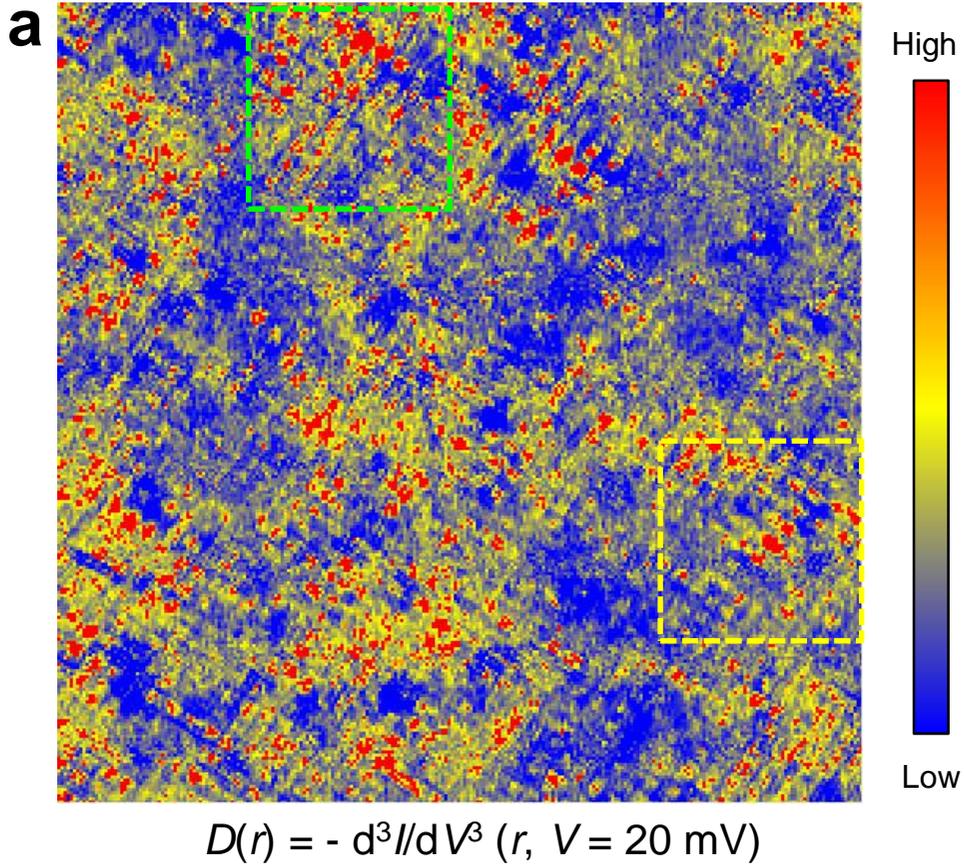

$D(r) = -\,d^3I/dV^3\,(r,\ V = 20\ \text{mV})$

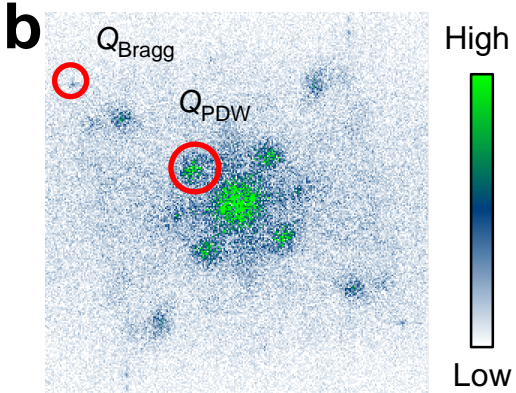

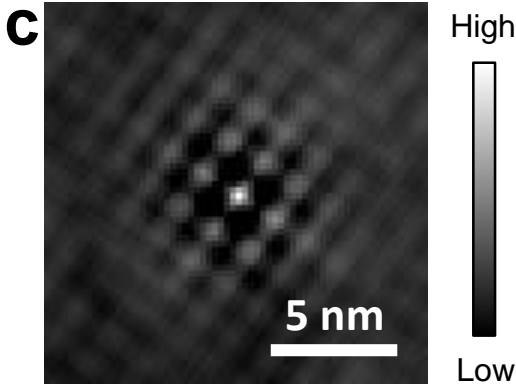

# Figure 4

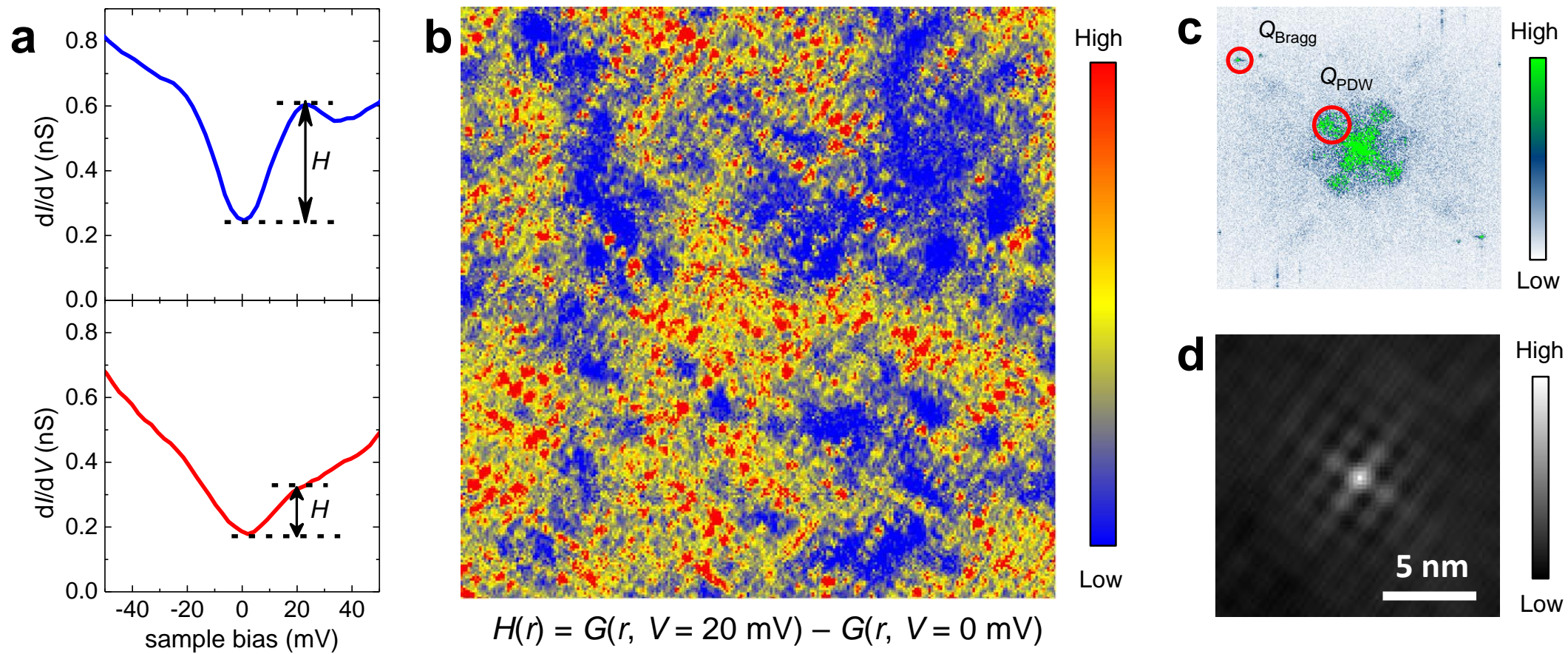

$H(r) = G(r, V = 20\ mV) - G(r, V = 0\ mV)$